\documentclass[prd,aps,twocolumn,superscriptaddress,floatfix,preprintnumbers]{revtex4}
\usepackage{amsmath}
\usepackage{graphicx}
\usepackage{xcolor}
\usepackage{amsfonts}
\usepackage{amssymb,amsmath,bm}
\usepackage{hyperref}
\usepackage[export]{adjustbox}
\hypersetup{colorlinks=true,citecolor=blue}

\usepackage{multirow}

\usepackage{physics}

\hypersetup{
    colorlinks=false,
    citecolor=blue,
    linkcolor=blue,
    linktoc=page
}

\newcommand{\avg}[1]{\ensuremath{\left\langle \,#1\, \right\rangle}}

\newcommand{\be}{\begin{equation}}
\newcommand{\ee}{\end{equation}}

\newcommand{\bea}{\begin{eqnarray}}
\newcommand{\eea}{\end{eqnarray}}
\newcommand{\bdm}{\begin{displaymath}}
\newcommand{\edm}{\end{displaymath}}

\newcommand{\bs}{\mathbf{s}}
\newcommand{\bk}{\mathbf{k}}
\newcommand{\bn}{\mathbf{n}}
\newcommand{\HH}{\mathcal{H}}
\newcommand{\ndv}{{v_{\parallel}}}

\begin{document}

\preprint{CERN-TH-2023-031}

\title{Large Scale Limit of the Observed Galaxy Power Spectrum}

\author{Matteo Foglieni}
\email{matteo.foglieni@lrz.de}
\affiliation{Leibniz Supercomputing Centre (LRZ), Boltzmannstraße 1, 85748 Garching bei München, Germany}
\affiliation{Dipartimento di Fisica ‘Aldo Pontremoli’, Universita’ degli Studi di Milano, \\ Via Celoria 16, 20133 Milan, Italy}
\author{Mattia Pantiri}
\email{mattia.pantiri@studenti.unimi.it}
\affiliation{Dipartimento di Fisica ‘Aldo Pontremoli’, Universita’ degli Studi di Milano, \\ Via Celoria 16, 20133 Milan, Italy}
\author{Enea Di Dio}
\email{enea.didio@cern.ch}
\affiliation{Theoretical Physics Department, CERN, 1211 Geneva 23, Switzerland}
\author{Emanuele Castorina}
\email{emanuele.castorina@unimi.it}
\affiliation{Dipartimento di Fisica ‘Aldo Pontremoli’, Universita’ degli Studi di Milano, \\ Via Celoria 16, 20133 Milan, Italy}

\begin{abstract}
The large scale limit of the galaxy power spectrum provides a unique window into the early Universe through a possible detection of scale dependent bias produced by primordial non Gaussianities. On such large scales, relativistic effects could become important and
be confused for a primordial signal. In this Letter we provide the first consistent estimate of such effects in the observed galaxy power spectrum, and discuss their possible degeneracy with local primordial non Gaussianities. We also clarify the physical differences between the two signatures, as revealed by their different sensitivity to the large scale gravitational potential.
Our results indicate that, while relativistic effects could easily account for 10\% of the observed power spectrum, the subset of those with a similar scale dependence to a primordial signal can be safely ignored for current galaxy surveys, but it will become relevant  for future observational programs. 
\end{abstract}
\maketitle

The distribution of the large scale structure (LSS) of the Universe encodes a wealth of information about the statistical properties of the primordial density fluctuations. In particular, the search for primordial non Gaussianities (PNG) in LSS data has become a central goal of current and future observational programs~\cite{Achucarro:2022qrl}, such as DESI~\cite{DESI:2016fyo}, Euclid~\cite{Amendola:2016saw}, SPHEREx~\cite{Dore:2014cca}, and Rubin Observatory~\cite{LSSTScience:2009jmu}. PNG could be generated by an inflationary phase in the early Universe, and any detection of PNG would have a significant impact on our understanding of cosmology. 

The LSS is primarily sensitive to the so-called local PNG, for which the non Gaussian signature is localized in the squeezed limit of the primordial curvature bispectrum. In single field, slow-roll, models of inflation, local PNG are, for all practical purposes, negligible \cite{Maldacena:2002vr,Alvarez:2014,Cabass:2016cgp}, therefore this entire class of models can be ruled out by a detection. Conversely, multifield models of inflation generically predict large local PNG \cite{Alvarez:2014,Achucarro:2022qrl}, and are thus increasingly constrained by a stronger upper bound. 

Local PNG are typically parametrized by a single number, $f_{\rm NL}$, defined via $\Phi_p (\vb{x}) = \phi_g(\vb{x}) + f_{\rm NL} \left[\phi_g(\vb{x})^2-\avg{\phi_g(\vb{x})^2} \right]$, where $\Phi_p$ is the primordial gravitational potential, and $\phi_g$ is a mean zero Gaussian random field. 
Measurements of the anisotropies of the cosmic microwave background (CMB) have put constraints on local PNG, $f_{\rm NL} = -0.8 \pm 5$~\cite{Planck:2019kim}, and this uncertainty is expected to shrink by another 50\% with upcoming CMB experiments \cite{CMB-S4:2016ple}. LSS data are still far from the theoretical benchmark of $f_{\rm NL}\sim \mathcal{O}(1)$, the most recent analyses reach $\sigma_{f_{\rm NL}}\sim 20\text{-}30$~\cite{Castorina2019,Mueller:2022dgf,Cabass:2022epm,DAmico:2022gki}, but future surveys are expected to improve these bounds by more than an order of magnitude \cite{Dore:2014cca,Sailer:2021yzm,Achucarro:2022qrl,Cabass:2022ymb,CosmicVisions21cm:2018rfq}. 
The advantage of LSS surveys over the CMB is that PNG leave a signature in the two-point correlation function of biased tracers like galaxies. Specifically, in the presence of local PNG, the bias relation between a discrete set of objects and the underlying density field $\delta_m$ is modified with respect to the one in a Gaussian Universe, becoming $\delta_g \sim b_1 \delta_m + f_{\rm NL} b_\phi \Phi_p $~\cite{Dalal:2008,Matarrese:2008,Slosar:2008hx,Desjacques:2010jw}, where $b_\phi$ is an $\mathcal{O}(1)$ number for  typical luminosity selected samples, although with large theoretical uncertainties \cite{Barreira:2021ueb,Barreira:2022sey}. Through Einstein's equations, the presence of the gravitational potential in the above expression implies that, on large scales, the power spectrum of galaxies acquires a distinct $k^{-2}$ signature. A detection of such scale dependence is then interpreted as a smoking gun of the presence of local PNG. Given the importance a measurement of nonzero PNG would have in shaping our understanding of the early Universe, it is of utmost importance to investigate whether these observational fingerprints are unique or can be mistaken for other physical processes.  

Concerning CMB observations, it was long ago realized that secondary effects, both at last scattering~\cite{Nitta2009,Creminelli:2011sq} and along the line of sight~\cite{Hu:2000ee,Smith:2006ud,Hanson2009,Mangilli2009,Lewis:2011fk,Hill:2018ypf,Coulton:2022wln}, produce a nonvanishing squeezed limit CMB bispectra, with an amplitude equivalent to $f_{\rm NL}\sim$ few. This bias had to be removed in the analysis of Planck data~\cite{Planck:2019kim}. 
On the other hand, the LSS community has been heavily debating for more than a decade whether the scale dependent bias induced by local PNG can be mimicked by other effects (see, e.g.,~\cite{Matarrese:2020why,dePutter:2015vga,Grimm:2020ays,Mitsou:2019ocs,Castorina:2021xzs,Cabass:2016cgp} for recent results). In particular, most of the attention is directed towards the so-called projection, or general relativistic (GR), effects introduced by the mapping between the observed coordinates of an object on the sky and its true position on our past light cone. Examples are the integrated Sachs-Wolfe (ISW) effect, weak lensing, or projections on the curved sky.

The goal of this Letter is to resolve this debate about the large scale limit of the observed power spectrum for any tracer of the LSS of the Universe. We think most of the confusion arises because computations are often not performed at the level of observables, but rather in terms of quantities and scales which are not the ones measured by the detectors and telescopes. Working at the level of the observed power spectrum of galaxies will resolve any possible ambiguities.

Our main result is that projection effects generate, on large scales, an observed power spectrum with a similar scale dependence to the one a local PNG signal with $b_\phi f_{\rm NL} \sim 1$ would produce. 
This, however, does not imply, as often conflated in the literature, that projection effects generate a coupling between the large scale gravitational potential and small scales physics. In \cite{Castorina:2021xzs} we have indeed shown that the observed power spectrum and correlation functions do not depend on the large scale value of the gravitational potential and its gradient: if we artificially move the observed patch of the sky into a constant gravitational field nothing will change, nor it will in a constant gradient. The assumptions that go into the above result are the same ones behind Maldacena's consistency relation~\cite{Maldacena:2002vr}, its extensions \cite{Creminelli:2012ed}, and the existence of Weinberg's adiabatic modes \cite{Weinberg:2003sw}, namely general relativity, the Gaussianity and adiabaticity of the initial conditions, and the absence of anisotropic stresses on large scales. Notice that the same conditions have been employed by the authors of \cite{dePutter:2015vga} to show the absence of scale dependent bias in a Gaussian Universe.
If local PNG are nonzero, then local physics responds to the presence of a long wavelength gravitational potential, through scale dependent bias and through super sample effects, which are nonzero even for dark matter \cite{Castorina2020,Castorina:2021xzs}. 

Local PNG and projection effects thus leave a similar imprint on large enough scales, but they remain physically distinct. Given a set of observations at a physical scale $s$, the latter are not sensitive to the gravitational potential with wavelength $ k^{-1} < s$, we say in the infrared (IR), while the former are, with an amplitude proportional to $f_{\rm NL}$. 
A consequence of the aforementioned result is that, in the absence of local PNG, no new additional parameter is required to model any $k^{-2}$-like terms, which are just a function of all the other cosmological and nuisance parameters of the model.
One of the goals of this Letter is to investigate how the amplitude of these terms depends on  the model parameters.

\textbf{The observed power spectrum--}
In order to determine if projection effects could mimic the signature of local PNG in LSS observables, we need to connect the observed power spectrum to the theoretical model that includes such terms. This  has been done in Ref.~\cite{Castorina:2021xzs}, of which we now summarize the main results. The starting point is the well-known expression for the observed number counts [see Eq.~\eqref{eq:number_counts}] \cite{Yoo:2009,Yoo:2010,Challinor:2011bk,Bonvin:2011bg}. Crucially, this expression contains fields evaluated at the observer's position, which are required by gauge invariance, and are instrumental for the cancellation of the IR divergences~\cite{Castorina:2021xzs,Ginat:2021nww,Mitsou:2023wes}. We then need an estimator for the three-dimensional power spectrum. In this Letter we take it to be the FKP estimator~\cite{FKP}
\bea
\hat P_0 \left( k \right) &=& 
A^{-1} \int \frac{\dd \Omega_{\hat\bk}}{4\pi} \int \dd ^3s_1 \dd ^3s_2 \,\Delta \left( \bs_1 \right) \Delta \left( \bs_2 \right) 
\nonumber \\
&&
\hspace{2cm}
W\left( \bs_1 \right) W\left( \bs_2 \right) e^{i \bk \cdot \left( \bs_1 - \bs_2 \right) } 
\eea
where $A$ is a normalization constant, $\Delta(\mathbf{s}_{1,2}) $ denotes the observed galaxy density contrast at position $\mathbf{s}_{1,2}$ and $W (\bs)$ is the survey window function. Other multipoles can be similarly measured, but in this Letter we focus only on the monopole, since it contains almost all the PNG signal if the other cosmological parameters are fixed \cite{Castorina2019}. 
The expectation value of the FKP estimator then reads
\be \label{eq:power_spectrum}
\langle \hat{P}_0 \left( k \right) \rangle = \sum_\ell\frac{1}{2\ell+1}  \int \dd s s^2\, \xi_\ell\left( s, z_{\rm eff } \right) j_0 \left( k s \right) Q_\ell \left( s \right)
\ee
where $\xi_\ell \left( s ,z_{\rm eff} \right)$ is the multipole expansion of the full-sky two-point correlation function, $\xi(s=|\bs_2-\bs_1|,s_1,\hat{s}\cdot\hat{s}_1)\equiv \avg{\Delta(\bs_1) \Delta(\bs_2)}$,  and the $Q_\ell$'s
are the multipoles of the two-point function of the survey mask \cite{Beutler:2018vpe}. In the above equation we made the further assumption that, despite the possible large extent in redshift of the survey, the redshift dependence of the clustering can be captured by evaluating the model at an effective redshift, $z_{\rm eff}$, defined by the radial selection function~\cite{Castorina:2021xzs} \footnote{We ignore the Integral Constraint (IC)\cite{deMattia:2019vdg}, since it does not qualitatively change the results. For our reference model, the IC is 5 to 10 times smaller than projection effects. In the modeling of the observed power spectrum we also did not include the contribution coming from the observer's velocity, which can be easily added \cite{Castorina:2021xzs,Elkhashab:2021lsk}.}. 

From Eq.~\eqref{eq:power_spectrum} we stress that the full-sky correlation function (see Appendix~\ref{app:number_counts} for more details) is the only physical ingredient needed to compute the observed power spectrum. 
The correlation function, and thus the observed power spectrum, is dominated, on all scales, by the terms proportional to the matter density field and the velocity gradients (redshift space distortions, RSD). In the remainder of this Letter we will refer to the latter as the Newtonian term.

By dimensional analysis of Eq.~\eqref{eq:number_counts}, contributions to the observed galaxy number counts proportional to the gravitational potential and to its time derivatives scale at least as $(\mathcal{H}/k)^2$ compared to the leading terms, where $\mathcal{H}$ is the conformal Hubble parameter, and could therefore contaminate a search for local PNG.
However, as anticipated in the introduction, the latter physically correspond to a correlation between small scale physics and the large scale gravitational potential, while the former are insensitive to the potentials in the IR of a given scale. 
This result, first proven in \cite{Castorina:2021xzs} (but see \cite{Scaccabarozzi:2018vux} for earlier results for the correlation function), relies on the existence of the consistency relations and of the Weinberg's adiabatic mode, and it holds only for the observed correlation function or power spectrum, not for the individual terms that contribute to the predictions.
For instance, the part of the observed correlation function proportional to two powers of the gravitational potential, $\avg{\Phi(\bs_1) \Phi(\bs_2)}$, diverges when more and more modes with wavelength $k\ll1/s$ are included in the calculation, but this divergence is canceled exactly by other contributions, independently of the value of the cosmological and nuisance parameters. We stress that the cancellation happens even in the absence of divergences, as this is purely a consequence of the decoupling between the small scale dynamics and the large scale potentials.

\begin{figure}[t!]
\centering
\includegraphics[width=0.475\textwidth]{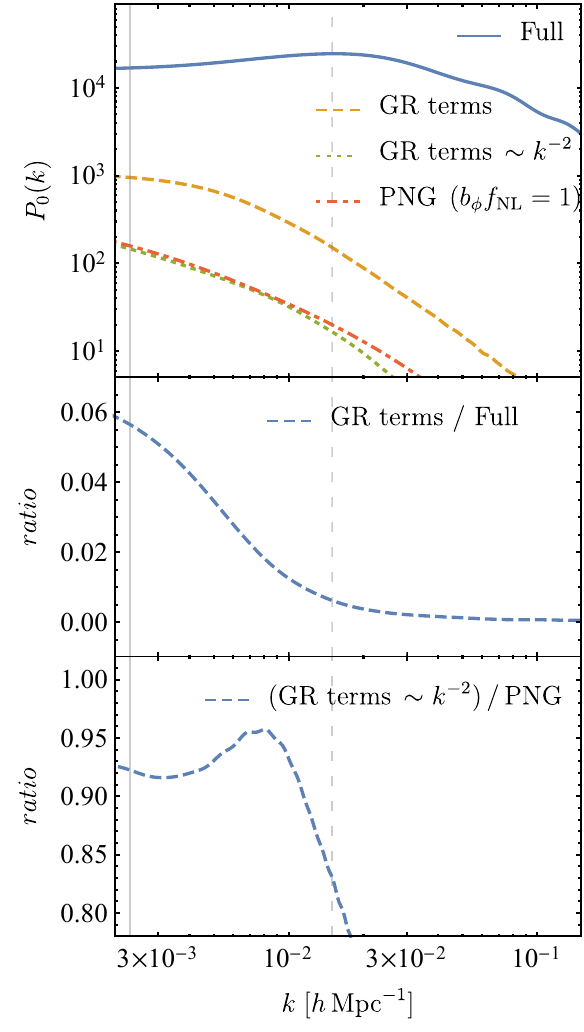}
\caption{Linear theory prediction for the observed galaxy power spectrum. We fixed $b_1=1.5$ and $f_{\rm evo} = s_b = 0$, and consider a survey between $z_{\rm min}=1$ and $z_{\rm max} = 1.5$ covering half of the sky. The two vertical lines correspond to the survey boundary and the equality scale, shown with continuous and dashed lines, respectively. \emph{Upper panel}: The monopole of the power spectrum $P_0(k)$ as a function of scale $k$. The continuous blue line shows the full result, the dashed orange one shows the sum of all the GR corrections, the dotted green line shows the subset of those with an approximate $k^{-2}$ scaling, and the dot-dashed red line corresponds to a $b_\phi f_{\rm NL} =1$ local PNG signal. \emph{Middle panel}: the ratio between the sum of all the projection effects and the full signal.
\emph{Lower panel:} The ratio between the sum of all the GR terms with a $k^{-2}$ scaling and the local PNG signal.}
\label{Fig:fnl}
\end{figure}
\textbf{Results and discussion--}
The last  missing ingredients in Eq.~\eqref{eq:power_spectrum} are the survey specifications, namely, the  angular footprint and radial selection function.
For simplicity, we assume our galaxy sample covers half of the sky from $z_{\rm min}=1.0$ to $z_{\rm max} = 1.5 $ \footnote{For an azimuthally symmetric window function, the convolution in Eq.~(\ref{eq:power_spectrum}) vastly simplifies~\cite{Castorina:2021xzs}.}. This corresponds to a volume of $V=34 ({\rm Gpc}/h)^3$ \footnote{The code offers the possibility to provide the multipoles of the window function for more realistic setups, like in DESI or Euclid, which are however not publicly available. Our choice thus allows the results of this letter to be reproduced and compared by other groups.}. We also need to specify the values of the galaxy bias $b_1$,  the evolution bias $f_{\rm evo}$ and the magnification bias $s_b$. Our fiducial model has $b_1 = 1.5$ and $f_{\rm evo} = s_b =0$, but we will also show results when the model parameters are allowed to vary.

All numerical results are obtained with the public code GaPSE, which is released together with this Letter~\footnote{A more detailed description of GaPSE will be presented in \cite{Pantiri23}. When possible, the code has been compared to the state-of-the art in the literature \cite{Tansella:2018sld,Jelic-Cizmek:2020jsn}.}\footnote{The code is available at \url{https://github.com/foglienimatteo/GaPSE.jl} .}. Unless otherwise noted, all numerical calculations assume a best-fit Planck 2018 cosmology~\cite{Aghanim:2018eyx}.

\begin{figure}[t]
\centering
\includegraphics[width=0.475\textwidth]{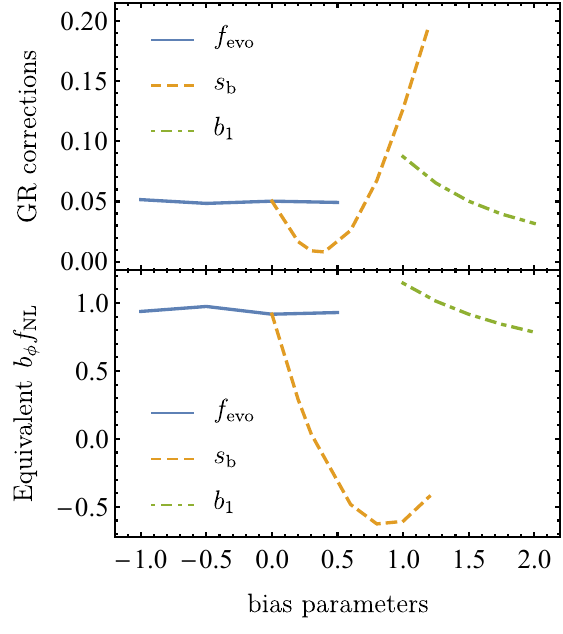}
\caption{\emph{Upper panel:} The average size of the contribution of projection effects to the observed monopole of the power spectrum, as a function of the model parameters. We vary one parameter at the time and keep the others fixed to the fiducial value in Fig.~\ref{Fig:fnl}. The blue line corresponds to changing only $f_{\rm evo}$, while the orange dashed and green dot-dashed ones to varying $s_b$ and $b_1$ respectively. \emph{Lower panel:} The equivalent amplitude of the local PNG signal, defined as the ratio of a $b_\phi f_{\rm NL}=1$  term to the GR contributions scaling as $k^{-2}$. For both plots the average is computed by    taking the mean of the corresponding quantity between $k= 0.004\,h$/Mpc and the survey boundary.}
\label{fig:eq_fnl}
\end{figure}

The upper panel of Fig.~\ref{Fig:fnl} shows the various terms contributing to the observed power spectrum. The blue line shows the full result, which is dominated by the Newtonian terms, while the orange dashed line shows the sum of all GR corrections, the latter dominated by lensing and wide-angle effects. 
The two vertical lines indicate the largest scale probed by our hypothetical survey and the matter-radiation equality scale~\footnote{The peak of the power spectrum is determined by the horizon scale at matter-radiation equality. Smaller scales that entered the horizon during the radiation-dominated epoch are suppressed relative to larger scales that entered the horizon during the matter-dominated epoch.}, respectively, with continuous and dashed gray lines.
We find that, for typical values of the bias parameters, the projection effects are below 10\% of the total signal in the monopole, much smaller than the expected cosmic variance for the volume considered.

To isolate the terms that could mimic a true local PNG signal we then remove from the total expression of the galaxy power spectrum the contributions proportional to the dark matter density field and its inverse gradients. These are all the auto- and cross-correlations between the Newtonian term and the lensing term, plus their cross-correlation with the Doppler term.
We show the remaining contributions as the green dotted line in the upper panel of Fig. \ref{Fig:fnl}. For comparison, we added, in dot-dashed red, the expected local PNG signal with $b_\phi f_{\rm NL} = 1$. 
Note that when subtracting the Newtonian terms, we include wide-angle effects, which can be easily calculated following \cite{Castorina:2017inr,Castorina2018,Castorina2019,Beutler:2018vpe}. This choice is motivated by the leakage, introduced by the survey mask \cite{Beutler:2018vpe}, of the dipole of the correlation function into the observed power spectrum monopole. This term scales as $k^{-1}$ and should be removed to isolate the contamination to a local PNG estimate. 

For our reference model with $s_b = f_{\rm evo} =0$ we see that, on large scales, projection effects have a similar shape to a $b_\phi f_{\rm NL}\sim 1$ signal. The degeneracy between the two is not perfect, due to the presence of the cross-correlation between the ISW effect and the Newtonian term, and the one between lensing and gravitational potentials. Both effects do not scale exactly as $k^{-2}$ as a consequence of the integral along the line of sight. 

It is worth stressing that our findings do not imply that projection effects should be ignored until a sensitivity to $f_{\rm NL}\sim 1$ is reached.
In an analysis of real data the importance of projection effects is estimated simultaneously with local PNG and other cosmological parameters, not by simply removing the contamination as discussed above. The final constraint will be the result of the complicated interplay
between the signal one is trying to measure and the uncertainty, i.e.,~the cosmic variance,
in the measurements themselves. If projection effects, like lensing or wide-angle corrections, can be measured with enough signal to noise, not including them in the modeling could actually result in a biased estimate of the amplitude of local PNG. A large positive value of $f_{\rm NL}$ could partially compensate, given the statistical uncertainties, for the missing terms in the theoretical model. A quantitative study of this problem would require the computation of the power spectrum covariance in the presence of GR effects. This could have a non-negligible effect on the final constraint on $f_{\rm NL}$, since the latter improves very quickly as larger and larger scales are included in the analysis, due to its supposedly unique dependence on wave numbers, which is now shared with projection effects. An exact implementation of the analytical power spectrum covariance is still missing in the literature, with approximations of the full result only available in the flat sky limit and on scales where the window function is subdominant \cite{Wadekar:2019rdu}.  Computing the exact analytical covariance in the presence of GR effects thus goes beyond the scope of this Letter, but we intend to return to it in forthcoming work. 
Tightly related to the estimate of the power spectrum covariance is the effect of observational systematic effects, like galactic foregrounds, which could severely limit, or even bias, the constraint on $f_{\rm NL}$, see e.g.,~Ref.~\cite{Rezaie:2021voi} for a recent investigation. A discussion of similar issues in projected galaxy clustering can be found in \cite{Camera:2014bwa, Alonso:2015sfa}.

It is also important to keep in mind that the equivalent value of local PNG induced by projection effects is sensitive to the volume probed by the survey, and to its angular and radial selection functions. While in this Letter we focused on a simple toy model for the window function, our public code GaPSE can handle arbitrary survey geometries. 
We conclude this section by investigating, for different values of $s_b$, $f_{\rm evo}$, and $b_1$, the amplitude of the GR contribution to the power spectrum monopole and the amplitude of the equivalent local PNG signal produced by projection effects.
The former is defined as the average, on scales between $k=0.004\,h$/Mpc and the survey boundary, of the GR corrections to the plane parallel expression, while the latter is the average, over the same scales, of the ratio between the approximate $k^{-2}$ GR contributions and the local PNG signal with $b_\phi f_{\rm NL} =1$. The result is presented in Fig. \ref{fig:eq_fnl}. The blue line corresponds to varying $f_{\rm evo}$ while keeping $s_b$ and $b_1$ at their fiducial value, varying $s_b$ is shown in orange, and varying $b_1$ in green. We see that the size of the projection effects is very insensitive to the value of $f_{\rm evo}$, and it decreases when the linear bias increases. This was expected since the Newtonian term scales like $b_1^2$. 
Projection effects are instead more sensitive to the value of $s_b$. For redshifts $z>1$, the autocorrelation of the lensing contribution quickly dominates over the other terms, and it scales approximately like $4 s_b^2$. However, a large value of magnification bias, $s_b>1$, is currently disfavored by the models of the luminosity function for DESI and Euclid galaxies \cite{Jelic-Cizmek:2020pkh,Elkhashab:2021lsk,Wang:2020ibf}. 

Similar conclusions apply to the equivalent PNG amplitude, which is only mildly dependent on $f_{\rm evo}$ and $b_1$, and it roughly corresponds to a $b_\phi f_{\rm NL} = 1$ signal. Also in this case we see a stronger dependence on the value of $s_b$, although it is still restricted to $|b_\phi f_{\rm NL}| \lesssim 1$. For large value of $s_b$, the cross-correlation between the ISW and the lensing term dominates the equivalent local PNG signal, implying the scaling is very much different than $k^{-2}$.  
Our findings are qualitatively in agreement with \cite{Jeong:2011as,Noorikuhani:2022bwc}, however a proper comparison is not possible since the estimates of the GR corrections in these works do not include a window function, the effect of the terms evaluated at the observer's position, and of several other possibly divergent terms. It cannot be emphasized enough that only by including all the contributions can we guarantee the robustness of the effective value of local PNG generated by projection effects.

\textbf{Conclusions and future directions--} In this Letter we studied the large scale limit of the observed galaxy power spectrum. We provided the first consistent comparison between a local PNG signal and projection effects for three-dimensional observables in Fourier space, and we robustly conclude that GR effects could produce an equivalent local PNG signal with $b_\phi f_{\rm NL} = 1$. We stress that our findings apply for any tracer of the LSS of the Universe, as long as it admits a perturbative bias expansion on large enough scales. 
Our results have implications for future observational programs aiming at  $\sigma(f_{\rm NL} )\lesssim1$. For these surveys, the accuracy at which we will be able to measure the amplitude of local PNG will depend on the uncertainty on $s_b$ and $f_{\rm evo}$, whose effect has to be included in order to not bias the constraints. In addition, projection effects will now contribute to the covariance, reducing the improvement in the determination of $f_{\rm NL}$ as the volume of the survey increases.

While the analysis in this Letter focused on the power spectrum, we expect our results to carry over to higher point statistics, e.g.~the bispectrum, for which the study of projection effects is still ongoing \cite{Yoo:2014sfa, Bertacca:2014dra, DiDio:2014lka, DiDio:2015bua, DiDio:2016gpd, Umeh:2016nuh, Bertacca:2017dzm, Koyama:2018ttg, Castiblanco:2018qsd, deWeerd:2019cae, Maartens:2020jzf, Magi:2022nfy, Montandon:2022ulz,Yoo:2022klz}. We plan to explore these directions in future work.

\vfill 

\textbf{Acknowledgements--} We thank Oliver Philcox for comments on an earlier version of the draft.
E.C. would like to thank Marko Simonovic for discussion on the squeezed limit of the CMB bispectrum. 

\vfill
\appendix
\section{The relativistic galaxy number counts}
\label{app:number_counts}
The correlation function can be easily computed from the expression for the fully relativistic number counts, as first studied to linear order in perturbation theory in Refs.~\cite{Yoo:2009,Yoo:2010,Challinor:2011bk,Bonvin:2011bg}.  Including also the observer terms, the galaxy number counts read as~\cite{Scaccabarozzi:2018vux,Grimm:2020ays,Castorina:2021xzs}
\begin{widetext}
\bea \label{eq:number_counts}
\Delta \left( \bn ,z \right) &=& \left\{b_1 D_m + \HH^{-1} \partial_r  \ndv   \right\}
+ \left\{\frac{5s_b-2}{2} \int_0^{r} \dd  r' \frac{r - r'}{r r'} \Delta_\Omega \left( \Psi + \Phi \right) \right\} 
+ \left\{ \mathcal{R} \left( \ndv - \ndv_o \right) { - \left( 2 - 5s_b\right) \ndv_o } \right\}
\nonumber \\
&&+ \left\{ \left( \mathcal{R} { - \frac{2-5s_b}{\HH_0 r}}\right) \HH_0 V_o + \left( \mathcal{R} +1 \right)\Psi -\mathcal{R} \Psi_o  
+ \left( 5 s_b -2 \right) \Phi  + \dot \Phi \HH^{-1} 
+ \left( f_{\rm evo} - 3 \right) \HH V \right\}
\nonumber \\
&&
+ \left\{ \frac{ 2 - 5 s_b }{ r} \int_0^{r} \left( \Psi + \Phi \right) \dd  r'
+ \mathcal{R}\int_0^{r} \left( \dot \Psi + \dot \Phi \right) \dd  r' \right\} \, ,
\eea
\end{widetext}
where we have introduced the coefficient
\be
\mathcal{R} = 5 s_b +\frac{2 - 5s_b}{\HH r} + \frac{\dot \HH} {\HH^2} - f_{\rm evo} \, ,
\ee
and three bias factors: the linear galaxy bias $b_1$, the magnification bias $s_b$, and the evolution bias $f_{\rm evo}$. 
We have grouped the different effects with curly brackets in the following order: standard density plus RSD, lensing convergence, Doppler, local gravitational potentials and integrated gravitational potentials. In the above expression, $D_m$ is the matter density fluctuation in the comoving gauge, $\Psi$ and $\Phi$ are the Bardeen's potentials, and $V$ is the velocity potential. Quantities evaluated at the observer's position carry the subscript $()_o$.

In full generality, the full-sky 2-point correlation function can be expressed as
\begin{widetext}
\be
\xi_{\mathcal{O}\mathcal{O'}} \left( s_1 , s_2 ,\hat \bs_1 \cdot \hat \bs_2 \right) = \int \frac{\dd  q}{2 \pi^2} q^2 P \left( q \right) \int_0^{s_1} \!\! \dd  \chi_1 \int_0^{s_2} \!\! \dd  \chi_2 
\mathcal{D}_{\mathcal{O}}\left( q, s_1 , \chi_1 \right) \mathcal{D}_{\mathcal{O'}}\left( q, s_2 , \chi_2 \right) j_0 \left( q \sqrt{\chi_1^2+ \chi_2^2 - 2 \chi_1 \chi_2 \hat \bs_1 \cdot \hat \bs_2} \right)
\ee
\end{widetext}
where $\mathcal{D}_{\mathcal{O}}$ is the differential operator associated to the perturbation $\mathcal{O}$ in eq.~\eqref{eq:number_counts}, through the dictionary provided in Ref.~\cite{Castorina:2021xzs}.

\bibliography{biblio_fnl_letter}

\end{document}